\documentclass[a4paper,11pt]{article}

\usepackage{contribution}
\newcommand{\weblink}[2][]{%
    \ifthenelse{\equal{#1}{}}%
    {\textnormal{\url{#2}}}%
    {\textnormal{\href{#2}{#1}}}%
}

\def\beq{\begin{equation}}
\def\eeq#1{\label{#1}\end{equation}}
\def\eeqn{\end{equation}}

\def\beqa{\begin{eqnarray}}
\def\eeqa#1{\label{#1}\end{eqnarray}}
\def\eeqan{\end{eqnarray}}

\let\bar=\overbar

\def\Dslash{\not{\hbox{\kern-4pt $D$}}}
\def\dslash{\not{\hbox{\kern-2pt $\del$}}}

\def\msb{{\bar{\ssstyle M \kern -1pt S}}}

\newcommand{\contribution}[7][]{%
  \clearpage
  \thispagestyle{plain}
  \ifthenelse{\equal{#1}{}}
  {\hypersetup{pdftitle={#2}}}
  {\hypersetup{pdftitle={#1}}}
  \hypersetup{pdfauthor={{#3} {#4}}}
  {\centering\normalfont\LARGE\bfseries\sffamily #2 \par\nobreak}
  \lhead{}
  \chead{%
    \textit{\footnotesize XIV International Conference on Hadron Spectroscopy
      (\weblink[\textit{hadron2011}]{http://www.hadron2011.de}), 13-17 June 2011, Munich, Germany}%
  }
  \rhead{}
  \bigskip
  \begin{center}
    {#3} {#4}\ifthenelse{\equal{#6}{}}{}{\footnote{\weblink[#6]{mailto:#6}}}
    \ifthenelse{\equal{#7}{}}{}{#7} \\
    \textit{#5}
  \end{center}
  \bigskip
}

\renewcommand{\abstract}[1]{%
  \begin{center}
    \begin{minipage}{0.85\textwidth}
      \begin{footnotesize}
        #1
      \end{footnotesize}
    \end{minipage}
  \end{center}
  \bigskip
}

\begin{document}

% % % % % % % % % % % % % % % % % % % % % % % % % % % % % % % % % % % % % % % % %
% your proceedings
%%%%%%%%%%%%%%%%%%%%%%%%%%%%%%%%%%%%%%%%%%%%%%%%%%%%%%%%%%%%%%%%%%%%%%%%%%%%%%%%%
%
% template for hadron2011 contribution
%
% please do not rename this file
%
% to create document run
%
%     pdflatex hadron2011.tex
%
%%%%%%%%%%%%%%%%%%%%%%%%%%%%%%%%%%%%%%%%%%%%%%%%%%%%%%%%%%%%%%%%%%%%%%%%%%%%%%%%%
{  % do not remove

%%%%%%%%%%%%%%%%%%%%%%%%%%%%%%%%%%%%%%%%%%%%%%%%%%%%%%%%%%%%%%%%%%%%%%%%%%%%%%%%%
% please define your macros here

\def \sd {\mathbf{D}}
\def \bs {\boldsymbol{\sigma}}
\def \bb {\mathbf{B}}
\def \be {\mathbf{E}}
\def \ba {\mathbf{A}}
\newcommand{\Tr}[1]{\textrm{Tr}\left[#1\right]}
\def \lvac {\langle0\vert}
\def \rvac {\vert0\rangle}
\def \br {\mathbf{r}}
\def \brg {\mathbf{R}}
\def \bx {\mathbf{x}}
\def \by {\mathbf{y}}
\def \bq {\mathbf{q}}
\def \aeff {\alpha_{\textrm{eff}}}
\def \cf {C_F}
\def \nc {N_c}
\def \ca {C_A}
\def \nf {n_f}
\def \tf {T_F}
\def \gmn {g^{\mu\nu}}
\def \gmnd {g_{\mu\nu}}
\def \id {\mathbf{1}}
\def \pmn {\Pi^{\mu\nu}}
\def \brh {\boldsymbol{\rho}}
\def \bl {\boldsymbol{\lambda}}
\def \bk {\mathbf{k}}
\def \mbk {\vert\bk\vert}
\def \bp {\mathbf{p}}
\def \bpg {\mathbf{P}}
\def \bsg {\mathbf{S}}
\newcommand{\Tint}[1]{{\hbox{$\sum$}\!\!\!\!\!\!\!\int\,}_{\!\!\!\!\raise-0.9ex\hbox{$\scriptstyle{#1}$}}}
\def \mq {\vert q^0\vert}
\def \mbq {\vert\bq\vert}
\def \lag {\mathcal{L}}
\def \qbar {\overline{q}}
\def \cc {\mathcal{C}}
\def\siml{{\ \lower-1.2pt\vbox{\hbox{\rlap{$<$}\lower6pt\vbox{\hbox{$\sim$}}}}\ }}
\def\simg{{\ \lower-1.2pt\vbox{\hbox{\rlap{$>$}\lower6pt\vbox{\hbox{$\sim$}}}}\ }}
\def \sd {\mathbf{D}}
\def \bs {\boldsymbol{\sigma}}
\def \bb {\mathbf{B}}
\def \be {\mathbf{E}}
\def \ba {\mathbf{A}}
\def \lvac{\langle0\vert}
\def \rvac {\vert0\rangle}
\def \br {\mathbf{r}}
\def \brg {\mathbf{R}}
\newcommand {\dt}[1] {\delta^{(3)}\left(#1\right)}
\def \bx {\mathbf{x}}
\def \by {\mathbf{y}}
\def \bq {\mathbf{q}}
\def \aeff {\alpha_{\textrm{eff}}}
\def \crr {C_R}
\def \gmn {g^{\mu\nu}}
\def \gmnd {g_{\mu\nu}}
\def \pmn {\Pi^{\mu\nu}}
\def \brh {\boldsymbol{\rho}}
\def \bl {\boldsymbol{\lambda}}
\def \bfnabla {\boldsymbol{\nabla}}
\def \bk {\mathbf{k}}
\def \mbk {\vert\bk\vert}
\def \bp {\mathbf{p}}
\def \bpg {\mathbf{P}}
\def \bsg {\mathbf{S}}
\def \mq {\vert q^0\vert}
\def \mbq {\vert\bq\vert}
\def \mbp {\vert\bp\vert}
\def \lag {\mathcal{L}}
\def \qbar {\overline{q}}
\def \cc {\mathcal{C}}
\def \trt {\tilde{\mathrm{Tr}}}
\def \als {\alpha_{\mathrm{s}}}
\def \csch {\text{csch}}
\def \m2   {\mu^{2 \epsilon}}
\def \phtl {\mathrm{pNRQCD}_\mathrm{HTL}}
\newcommand{\order}[1]{\mathcal{O}\left(#1\right)}
\newcommand{\MS}{{\overline{\rm MS}}}
\def\alVs{\alpha_{V_s}}
\def\alVo{\alpha_{V_o}}
\def\siml{{\ \lower-1.2pt\vbox{\hbox{\rlap{$<$}\lower6pt\vbox{\hbox{$\sim$}}}}\ }}
\def\simg{{\ \lower-1.2pt\vbox{\hbox{\rlap{$>$}\lower6pt\vbox{\hbox{$\sim$}}}}\ }}
\def\lqcd{\Lambda_{\mathrm{QCD}}}
\def\nn {\nonumber}
\def\bfsigma{\mbox{\boldmath $\sigma$}}
\def\sgn{\mathrm{sgn}}

%
%%%%%%%%%%%%%%%%%%%%%%%%%%%%%%%%%%%%%%%%%%%%%%%%%%%%%%%%%%%%%%%%%%%%%%%%%%%%%%%%%

%%%%%%%%%%%%%%%%%%%%%%%%%%%%%%%%%%%%%%%%%%%%%%%%%%%%%%%%%%%%%%%%%%%%%%%%%%%%%%%%%
% define title, author, and address
% contribution[short title]{title}{author first name}{author last name}{author address}{author email}{collaboration}
% the short title will appear in the page headers and the TOC of the book of proceedings
% the last two arguments may be left empty
\contribution % short title (optional)
{Heavy quarkonium spectrum and width in a weakly-coupled quark-gluon plasma}  % title
{Jacopo}{Ghiglieri}  % first and last name of author
{ Physik-Department, Technische Universit\"at M\"unchen,\\
James-Franck-Str. 1, 85748 Garching, Germany and Excellence Cluster Universe, Technische Universit\"at M\"unchen, Boltzmannstr. 2, 85748, Garching, Germany}  % author address
{jacopo.ghiglieri@ph.tum.de}  % author email optional
{}%
%%%%%%%%%%%%%%%%%%%%%%%%%%%%%%%%%%%%%%%%%%%%%%%%%%%%%%%%%%%%%%%%%%%%%%%%%%%%%%%%%

%%%%%%%%%%%%%%%%%%%%%%%%%%%%%%%%%%%%%%%%%%%%%%%%%%%%%%%%%%%%%%%%%%%%%%%%%%%%%%%%%
% abstract
\abstract{%
	We report a recent  calculation of the heavy quarkonium energy levels and decay widths in a quark- gluon plasma whose temperature is much smaller than the inverse radius of the bound state, based on a Non-Relativistic Effective Field theory framework for heavy quarkonium at finite temperature. Relevance for the phenomenology of the $\Upsilon(1S)$ in heavy ion collisions is also discussed.
}
%
%%%%%%%%%%%%%%%%%%%%%%%%%%%%%%%%%%%%%%%%%%%%%%%%%%%%%%%%%%%%%%%%%%%%%%%%%%%%%%%%%
\section{Introduction}
The suppression of quarkonium has been hypothesized 25 years ago \cite{Matsui:1986dk} to represent a signature of the formation of a deconfined medium and has been ever since intensely investigated, both theoretically and experimentally. Here we address the problem, central to these studies, of the behaviour of a quarkonium bound state in a deconfined thermal medium. To this end we shall illustrate the Effective Field Theory (EFT) framework that has been recently constructed in \cite{Brambilla:2008cx,Brambilla:2010vq,Brambilla:2011mk} (see also \cite{Escobedo:2008sy,Escobedo:2010tu,Escobedo:2011ie} for an analogous EFT of QED) by generalizing the successful zero-temperature framework of Non-Relativistic (NR) EFTs for heavy quarkonia to finite temperatures. These NR EFTs exploit the hierarchy $m\gg mv\gg mv^2$ that characterizes any NR binary bound state, $m$ being in this case the heavy quark mass and $v$ the relative velocity. $mv$ is then the typical momentum transfer or inverse radius and $mv^2\sim E$ the binding energy. The low-lying quarkonium states, especially the bottomonium ground states $\Upsilon(1S)$ and $\eta_b$, are believed to be approximately Coulombic. That corresponds to having $mv\sim m\als\gg\Lambda_{\mathrm{QCD}}$ and $mv^2\sim m\als^2\simg \lqcd$.\\
In a weakly-coupled plasma, which we consider in our study, the temperature $T$ and the chromoelectric screening mass $m_D$ are larger than $\lqcd$ and, since $m^2_D\sim g^2 T^2$, $T\gg m_D$. Under these conditions one can then calculate observables relevant for the phenomenology of low-lying states to a large extent analytically in perturbation theory, which makes them extremely interesting, also in the light of the recent CMS measurements of the suppression of the $\Upsilon$ family \cite{Chatrchyan:2011pe}.\\ 
In \cite{Laine:2006ns} the perturbative $Q\overline{Q}$ static potential was computed in QCD for distances $r$ such that $T\gg1/r\simg m_D$. The resulting potential surprisingly shows an imaginary part which is larger than the screened real part for $1/r\sim m_D$. This imaginary part can be traced back to the imaginary part of the gluon self-energy and is due to the Landau-damping phenomenon; it eventually leads to a thermal width for the bound state, which is in turn responsible for its dissociation, representing a change from the previous colour-screening paradigm. This change was further reinforced by the introduction of a \emph{dissociation temperature} in \cite{Escobedo:2008sy,Laine:2008cf}, defined as the temperature for which the imaginary part of the potentials becomes of the same size of its real part; parametrically it is of order $m\als^{2/3}$ and hence smaller than the temperature at which screening sets in. A quantitative calculation of the dissociation temperature for the $\Upsilon(1S)$ can be found in \cite{Escobedo:2010tu} and a phenomenological analysis, based on these imaginary parts, of $b\overline{b}$ bound states at LHC energies can be found in \cite{Strickland:2011mw}.\\
In \cite{Brambilla:2008cx} the static $Q\overline{Q}$ was first studied in an EFT framework, systematically exploring the hierarchy of different energy scales in the problem. Many possibilities were considered, from temperature smaller than $E$ to temperatures much larger than $1/r$, where the results of \cite{Laine:2006ns} were recovered in a rigorous EFT derivation. Furthermore a new dissociation mechanism, the colour-singlet to octet decay, was identified; it is the leading one when $E\gg m_D$. In \cite{Brambilla:2010xn} the relation between the proper real-time quarkonium potential and the correlator of two Polyakov loops, a quantity often measured on the lattice and used as input for potential models, was investigated. The breaking of Lorentz invariance induced by the preferred reference frame introduced by the medium was instead analyzed in \cite{Brambilla:2011mk} in the spin-orbit sector of the EFT. In the following we will report about the findings of \cite{Brambilla:2010vq}, where in a specific range of temperatures  the spectrum and width of quarkonia have been computed up to order $m\als^5$. To this end, the specific global hierarchy that the NR and thermodynamical scales fulfill in the assumed range of parameters has been exploited by constructing a corresponding tower of EFTs.
\section{Energy scales, the EFT formalism and the results}
The aforementioned global hierarchy we assume is $m\gg m\als\gg \pi T\gg m\als^2\gg m_D$,
which implies a temperature below the dissociation temperature. We also remark that for the $\Upsilon(1S)$ at the LHC it may hold that $m_b\approx 5\;\text{GeV}>m_b\als\approx1.5\;\text{GeV}>\pi T\approx 1\;\text{GeV}>m\als^2\approx 0.5\;\text{GeV}\simg m_D$.\\
Given this hierarchy, we now proceed to integrate out each scale in sequence. The integration of the mass scale $m$ yields non-relativistic QCD (NRQCD) \cite{Caswell:1985ui} and the further integration of the scale $m\als$ from NRQCD gives potential non-relativistic QCD (pNRQCD) \cite{Pineda:1997bj}. Since the temperature is much smaller than these two scales, it may be set to zero in the matching and the Lagrangians of NRQCD and pNRQCD are the same as at zero temperature.\\
Integrating out $T$ from pNRQCD modifies it into its Hard Thermal Loop (HTL) version, pNRQCD${}_{\mathrm{HTL}}$ \cite{Brambilla:2008cx,Vairo:2009ih}, where the light degrees of freedom are described by the HTL effective Lagrangian \cite{Braaten:1991gm} and the pNRQCD potentials receive a thermal part. Finally, within this EFT we can compute contribution to the spectrum and width from the scales $E$ and $m_D$. Diagrams contributing to the calculation are shown in Fig.~\ref{figure}.\begin{figure}[ht]
	\begin{center}
		\includegraphics[scale=0.4]{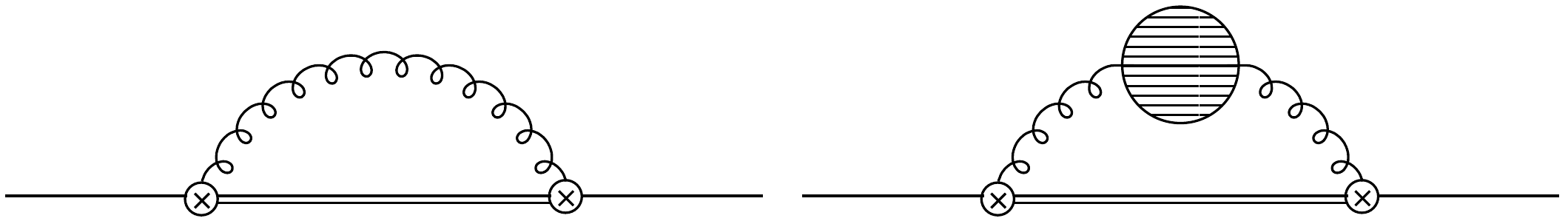}
	\end{center}
	\caption{The diagrams contributing to our calculation. Single lines are colour-singlet $Q\overline{Q}$ states, double lines are colour octets, curly lines are gluons, vertices are chromoelectric dipoles and the blob is the gluon self-energy. The imaginary part of the first diagram yields the singlet-to-octet decay mechanism, whereas the second one gives the Landau damping contribution to the width.}
	\label{figure}
\end{figure}\\
Let us now show the final results for the thermal contribution to the spectrum and to the width up to order $m\als^5$. We recall that for a Coulombic bound state the spectrum is at LO given by the Bohr levels $E_n=-m\cf^2\als^2/(4n^2)$ and the Bohr radius is $a_0=2/(m\cf\als)$. The vacuum contribution to the spectrum up to order $m\als^5$ can be read from \cite{Brambilla:1999xj}.\\
The thermal contribution to the spectrum then reads
\begin{eqnarray}
\nonumber
\delta E_{n,l}^{(\mathrm{thermal})}&=&
\frac{\pi}{9}N_c C_F \,\als^2 \,T^2 \frac{a_0}{2}\left[3n^2-l(l+1)\right] +\frac{\pi}{3}C_F^2\, \als^2\, T^2\,a_0
\\
\nonumber&&
+\frac{E_n\als^3}{3\pi}\left[\log\left(\frac{2\pi T}{E_1}\right)^2-2\gamma_E\right]
\left\{\frac{4 C_F^3\delta_{l0} }{n}+\frac{2N_c^2C_F}{n(2l+1)} +\frac{N_c^3}{4}
\right.
\\
\nonumber 
&& \hspace{3.2cm}
\left. +N_c \cf^2\left[\frac{8}{n (2l+1)}-\frac{1}{n^2} 
- \frac{2\delta_{l0}}{ n }   \right]\right\}
+\frac{2E_n\cf^3\als^3}{3\pi}L_{n,l}
\\
\nonumber
&&
+ \frac{a_0^2n^2}{2}\left[5n^2+1-3l(l+1)\right]
\left\{- \left[\frac{3}{2\pi} \zeta(3)+\frac{\pi}{3}\right]  C_F \, \als  \, T \,m_D^2
\right.
\\
&& \hspace{6.7cm}
+ \left. \frac{2}{3} \zeta(3)\, N_c C_F \, \als^2 \, T^3\right\},
\label{finalspectrum}
\end{eqnarray}
where $L_{n,l}$ is the QCD Bethe log \cite{Brambilla:1999xj}. The terms on the first line are the leading ones the power counting of the EFT and, being positive, lead to an increase in the mass of the bound state quadratic with the temperature.\\
For what concerns the thermal width, we have
\begin{eqnarray}
\nonumber
\Gamma_{n,l}^{(\mathrm{thermal})}&=&
\frac{1}{3}N_c^2C_F\als^3T+\frac{4}{3}\frac{C_F^2\als^3 T}{n^2}(\cf+\nc)
\\
\nonumber&&
+\frac{2E_n\als^3}{3}\left\{\frac{4 C_F^3\delta_{l0} }{n}+N_c \cf^2
\left[\frac{8}{n (2l+1)}-\frac{1}{n^2} - \frac{2\delta_{l0}}{ n }   \right]
+\frac{2N_c^2C_F}{n(2l+1)}+\frac{N_c^3}{4}\right\}
\\
\nonumber&&
-a_0^2n^2\left[5n^2+1-3l(l+1)\right]\left[
\left(\ln\frac{E_1^2}{T^2}+ 2\gamma_E -3 -\log 4- 2 \frac{\zeta^\prime(2)}{\zeta(2)} \right)\right.\\
&& \hspace{2cm}\left.\times\frac{C_F}{6} \als T m_D^2+\frac{4\pi}{9} \ln 2 \; N_c C_F \,  \als^2\, T^3 \right]
+\frac{8}{3}\cf\als\, Tm_D^2\,a_0^2n^4
\,I_{n,l}\;,
\label{finalwidth}
\end{eqnarray}
where $I_{n,l}$ is a new Bethe logarithm \cite{Brambilla:2010vq}. The terms in the first two lines are the leading ones and are caused by singlet-to-octet decay, whereas those on the last two lines are due to Landau damping. The width is at leading order linear in the temperature and much smaller than the binding energy. This small width is certainly not in contradiction with the recent CMS results \cite{Chatrchyan:2011pe} that point to a substantial survival of the $\Upsilon(1S)$ at the LHC.

%%%%%%%%%%%%%%%%%%%%%%%%%%%%%%%%%%%%%%%%%%%%%%%%%%%%%%%%%%%%%%%%%%%%%%%%%%%%%%%%%
% main text
% for short contributions sections are optional

%%%%%%%%%%%%%%%%%%%%%%%%%%%%%%%%%%%%%%%%%%%%%%%%%%%%%%%%%%%%%%%%%%%%%%%%%%%%%%%%%
% bibliographic items can be constructed using the LaTeX format in SPIRES
% see http://www.slac.stanford.edu/spires/hep/latex.html
% SPIRES will also supply the CITATION line information; please include it

%
%%%%%%%%%%%%%%%%%%%%%%%%%%%%%%%%%%%%%%%%%%%%%%%%%%%%%%%%%%%%%%%%%%%%%%%%%%%%%%%%%

}  % do not remove

%%% Local Variables: 
%%% mode: latex
%%% TeX-master: "../hadron2011.tex"
%%% End: 

\end{document}